# Automatic Classification of White Blood Cell Images using Convolutional Neural Network

Rabia Asghar, Arslan Shaukat, Usman Akram, Rimsha Tariq

National University of Sciences and Technology (NUST), Islamabad, Pakistan
rabiaasghar82@yahoo.com, arslanshaukat@ceme.nust.edu.pk,
usmakram@gmail.com, rimsha.ramo@gmail.com

**Abstract.** Human immune system contains white blood cells (WBC) that are good indicator of many diseases like bacterial infections, AIDS, cancer, spleen, etc. White blood cells have been sub classified into four types: monocytes, lymphocytes, eosinophils and neutrophils on the basis of their nucleus, shape and cytoplasm. Traditionally in laboratories, pathologists and hematologists analyze these blood cells through microscope and then classify them manually. This manual process takes more time and increases the chance of human error. Hence, there is a need to automate this process. In this paper, first we have used different CNN pre-train models such as ResNet-50, InceptionV3, VGG16 and MobileNetV2 to automatically classify the white blood cells. These pre-train models are applied on Kaggle dataset of microscopic images. Although we achieved reasonable accuracy ranging between 92 to 95%, still there is need to enhance the performance. Hence, inspired by these architectures, a framework has been proposed to automatically categorize the four kinds of white blood cells with increased accuracy. The aim is to develop a convolution neural network (CNN) based classification system with decent generalization ability. The proposed CNN model has been tested on white blood cells images from Kaggle and LISC datasets. Accuracy achieved is 99.57% and 98.67% for both datasets respectively. Our proposed convolutional neural network-based model provides competitive performance as compared to previous results reported in literature.

## 1 Introduction

White Blood Cells (WBCs) are an essential part of our immune system, defending our body against germs by ingesting them or producing antibodies. White blood cells are grouped into four different types: a) Monocytes (3-9%), b) Lymphocytes (20-40%), c) Eosinophils (2-4%), d) Neutrophils (50-70%). The percentage values are the standard range of WBCs present in the blood of a normal healthy person. An absence or disparity in the number of any white blood cells type can be caused by different diseases [1]. These diseases are identified by examining blood samples. Each subtype of WBC helps in tackling certain types of disorders such as cancer, hepatitis, HIV, allergic infections nephrotic syndrome, etc. It is necessary to count and recognize the human's WBCs to identify these diseases. Conventionally, hematology specialists perform categorization and numbering of these WBCs manually with the assistance of a microscope. However, this procedure is time exhausting, error sensitive, and complicated to operate.

The growth of image processing and artificial intelligence in the field of medicine has released physicians from the burden of manual categorization. Many researchers have proposed different methods to classify WBCs efficiently. Some of them used texture and geometrical feature classification methods, which includes preprocessing stage, where medical images are denoised and region of interest is segmented out [2], [3]. The next stage is feature extraction, where different texture and statistical features are obtained from an image and then passed to some machine learning classification algorithms such as K nearest neighbor, Support vector machines, and Neural networks. Another way is using image pixel as features and passing them through a convolution neural network (CNN), which consists of convolutions layers, fully connected layers and an output layer [4]. Researchers have used both classical machine learning algorithms as well as CNN models for WBCs classification [6].

Roysadi et. al. [5] presented a technique, segmenting WBCs and recognizing them using K-means algorithm. Gautam et. al. [7] also used feature-based approach for classifying WBCs. For this purpose, they used Naïve Bayes classifier and obtained 80.88% accuracy for this problem. Y. Wei et. al. [8] performed recognition of 7 types of white blood cells using CNN. They achieved accuracy of 88.5% using this method. P. Tiwariet.al. [10] did their research on the same problem, where their CNN architecture included two convolutional layers, a single pooling layer, one fully connected hidden layer followed by an output layer. They achieved 94% accuracy for two classes and 78% accuracy for four classes. H. Fan et. al. [12] proposed a method of end-to-end leukocyte localization and segmentation named as 'Leukocyte-Mask'. M. Z. Othman et. al. [11] proposed to use MLP-BP multilayer perceptron backpropagation for classification of white blood cells. 96% of classification accuracy was achieved in their study. A. Cinar et. al. [20] performed classification of four types of white blood cells using Alexnet Googlenet-SVM. Using this architecture, they achieved an accuracy of 99.73% for Kaggle dataset and 98.23% for LISC dataset.

From the literature, it is noted that the classification of white blood cells is quite popular. A large amount of work has been done focusing on image classification and segmentation. Few researchers have preferred handcrafted features for classification purposes. The earlier classification methods of white blood cells consist of pre-processing, feature selection and feature extraction steps. There is a recent demand to use convolutional neural networks (CNN) to enhance the performance of the classification systems of various white blood cells.

The aim of this research is to develop a model for the classification of various white blood cells, using machine learning methods. For this purpose, firstly, we have tested the CNN pre-trained models: VGG-16, ResNet50, InceptionV3, and MobileNetV2 for microscopic image dataset provided by Kaggle and evaluated their performance. Then we have introduced a CNN model for the classification of four major WBCs subtypes. The datasets used for this research are Kaggle and LISC. Our work aims to develop a convolution neural network (CNN) based model with decent generalization ability for the classification of various types of WBCs.

The paper is organized as follows. Section 2 presents the methodology and proposed architecture. Section 3 reports the results and their analysis with discussion. The last section draws the conclusions.

## 2 Proposed Methodology

This section illustrates the methodology undertaken to classify white blood cells.

### 2.1 White Blood Cell Datasets

The datasets used for this research are obtained from Kaggle and LISC.

**Kaggle Dataset.** The dataset contains 12,500 JPEG images of size 320 x 240 [18]. Each class has approximately 3000 images. The dataset has four different categories of WBC's; Eosinophils, Lymphocytes, Monocytes and Neutrophils, and images of each class are shown in Figure 1.

**LISC Dataset.** In this dataset, images are obtained from peripheral blood of 8 normal subjects, and 400 samples are taken from 100 microscope slides saved in BMP format [19]. Rotation augmentation method is applied to these images with degree of rotation set to 90, 180, 270 degrees. After augmentation, the LISC dataset contains 10,000 images of 720 x 576 size. The dataset also targets four different types of white blood cells similar to Kaggle dataset. The sample image of each type can be seen in Figure 2.

For training our model, all images in both datasets are resized to 100 x 100 and divided into three portions as 70%, 20% and 10%. The major part of both datasets is used for training, while other two portions are used for validation and testing.

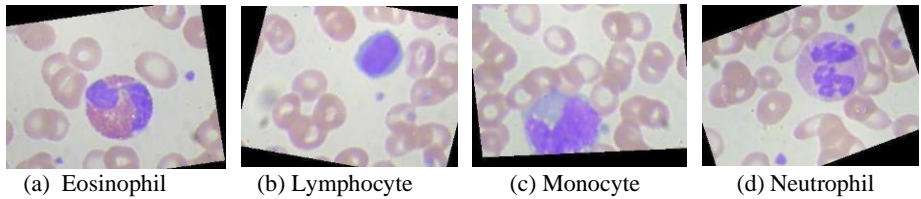

(a) Eosinophil  (b) Lymphocyte  (c) Monocyte  (d) Neutrophil

**Fig. 1.** Sample images in Kaggle dataset

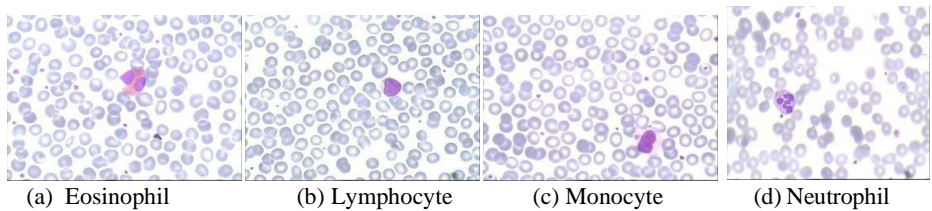

(a) Eosinophil  (b) Lymphocyte  (c) Monocyte  (d) Neutrophil

**Fig. 2.** Sample images in LISC dataset

### 2.2 Pre-trained CNN Models

Initially, we have used different pre-trained CNN models, ResNet-50, InceptionV3, VGG16 and MobileNetV2 for white blood cells classification. The details of these models are mentioned next.

**VGG16.** VGG16 is a pre-trained CNN architecture which contains 13 convolution layers, with pooling, and three fully connected hidden and output layers. The final output layer is the soft-max activation layer.

**ResNet50.** ResNet50 (Residual Neural Network) [22] pre-trained architecture contains total five stages. Each stage consists of convolutional layers and identity block. After average pooling layers, it ends with a fully connected output layer containing 4 nodes.

**InceptionV3.** Inception-V3 [21] connects convolutional layers through multilayer perceptrons that can learn non-linear functions. It has various symmetric and a-symmetric building chunks, including convolution layers, maximum and average pooling layers, dropout layers, concatenated and fully connected (FC) layers.

**MobileNetV2.** In MobileNetV2 [23], there are 2 different types of chunks. Both types of blocks consist of three layers. Moreover, first layer is convolution with rectified linear unit (Relu). The second layer is a depth-wise convolutional layer. And the third layer is 1×1 convolutional layer without non linearity.

### 2.3 Proposed Convolutional Neural Network

After testing the four pre-trained models on Kaggle dataset, it is observed that performance of these models is relatively low for white blood cell classification. Inspired by these four architectures and to enhance the accuracy, we have proposed our own convolutional neural network (CNN) architecture shown in Figure3. It consists of three convolutional layers, three pooling layers, two fully connected hidden layers and an output layer.

**Convolutional Layer.** The key layer in CNN is convolutional layer. This layer extracts a features-maps from an image. As highlighted before, our architecture contains three convolutional layers listed below:

1) First Layer: Kernel size: 3 x 3, number of filters: 32, Activation function: Relu, Stride: 1 and input size: 100 x 100 (3 channels).
2) Second Layer: Kernel size: 3 x 3, number of filters: 64, Activation function: Relu, Stride: 2.
3) Third Layer: Kernal size: 3 x 3, number of filters: 64, Activation function: Relu, Stride: 1

**Pooling Layer.** Pooling layer gradually reduces the resolution and size of the input image. We used max pooling with same parameters for all three pooling layers in our architecture with details given: Pooling type: Maximum, Pooling Size: 2 x 2, Stride: 1, Dropout 0.20.

**Fully Connected Layer.** The last layers of CNN model are fully connected layers. In our proposed methodology, we have two fully connected hidden layers and one fully connected output layer listed below:

1) Fully Connected Hidden Layer: Total nodes: 64, Activation: Relu.

2) Fully Connected Output Layer: Total nodes: number of classes, Activation: Softmax.

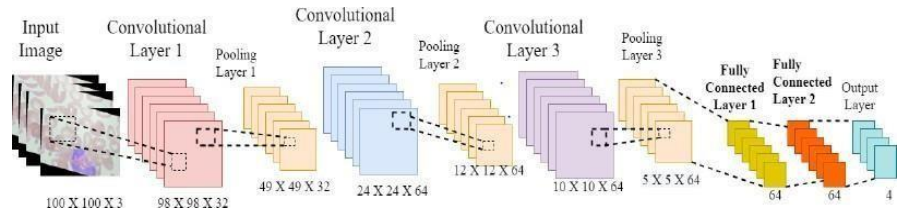

**Fig.3.** Architecture of our proposed CNN Model

## 3 Experimentation, Results and Discussion

In this section, we report the experimental results and present our detailed analysis and discussion.

### 3.1 Evaluation Parameters
There are certain parameters used in machine learning to determine the performance of the model. This work has used four generally utilized parameters: accuracy, recall, precision, and F-measure to evaluate the model fitness.

### 3.2 Pre-Trained Model Compilation and Results
Firstly, pre-trained architectures, VGG16, ResNet50, InceptionV3 and Mobile NetV2 are evaluated for Kaggle dataset. These architectures have been trained using sparse categorical cross entropy loss with Adam gradient-based optimized. The pre-trained weights used are for ImageNet dataset classification. The model is trained for last dense layers for 150 epochs with the learning rate of 0.0001.

Performance parameter values obtained with pre-trained architectures are presented in Table 1 on Kaggle dataset. With VGG-16, 15 images from eosinophils, 11 from lymphocytes, 14 images from monocytes and 40 from neutrophils, 80 images in total of 992 have been misclassified. Overall accuracy of 92% is achieved with VGG16. A total of 16 images from eosinophils, 17 from lymphocytes, 8 images from monocytes and 38 from neutrophils, 79 images are misclassified by using ResNet50. Overall accuracy of 92.3% is obtained. Overall, 95.67% accuracy is achieved with InceptionV3 model, where comparatively better results are obtained. With this pre-trained architecture, 15 images from eosinophils, 4 from lymphocytes, 7 from monocytes and 18 images from neutrophils are classified incorrectly. In the WBC's classification using MobileNetV2, 29 images from eosinophils, 7 from lymphocytes, 11 images from monocytes and 17 images from neutrophils are misclassified. 93.6% accuracy is achieved.

### 3.3 Proposed CNN Model Results
In quest for better performance and inspired by these architectures, we have proposed and trained our own CNN model. For training, three major parameters are decided, loss function, optimizer, and metrics of evaluation. Our CNN model uses loss function of sparse categorical cross entropy with default Adam gradient-based optimizer. The training dataset is fed to our model and is trained for 150 epochs, saving best weights depending on loss function. The proposed convolutional neural network model has been tested with WBC's images from both Kaggle and LISC datasets.

Table 1. Test results of pre-trained models on Kaggle dataset

| Model | Type | Truth | Classified | Accuracy (%) | Precision (%) | Recall (%) | F-measure (%) |
|---|---|---|---|---|---|---|---|
| VGG16 | Eosinophils | 233 | 248 | 0.930 | 0.93 | 0.82 | 0.87 |
|  | Lymphocytes | 237 | 248 | 0.961 | 0.96 | 0.97 | 0.96 |
|  | Monocytes | 234 | 248 | 0.950 | 0.95 | 0.98 | 0.96 |
|  | Neutrophils | 208 | 248 | 0.841 | 0.84 | 0.91 | 0.88 |
| ResNet50 | Eosinophils | 232 | 248 | 0.94 | 0.94 | 0.99 | 0.96 |
|  | Lymphocytes | 231 | 248 | 0.92 | 0.92 | 0.80 | 0.86 |
|  | Monocytes | 240 | 248 | 0.97 | 0.97 | 0.99 | 0.98 |
|  | Neutrophils | 210 | 248 | 0.86 | 0.86 | 0.90 | 0.88 |
| InceptionV3 | Eosinophils | 233 | 248 | 0.942 | 0.94 | 0.92 | 0.93 |
|  | Lymphocytes | 244 | 248 | 0.982 | 0.98 | 0.99 | 0.98 |
|  | Monocytes | 241 | 248 | 0.972 | 0.97 | 0.99 | 0.98 |
|  | Neutrophils | 230 | 248 | 0.932 | 0.93 | 0.93 | 0.93 |
| MobileNetV2 | Eosinophils | 219 | 248 | 0.890 | 0.89 | 0.92 | 0.90 |
|  | Lymphocytes | 241 | 248 | 0.971 | 0.97 | 0.97 | 0.97 |
|  | Monocytes | 237 | 248 | 0.952 | 0.95 | 0.98 | 0.97 |
|  | Neutrophils | 231 | 248 | 0.932 | 0.93 | 0.88 | 0.90 |

**Results on Kaggle Dataset.** Graph for model loss of Kaggle dataset and accuracy achieved along each epoch is shown in Figure 4. The performance of network is measured through cross entropy loss function, which is widely used to evaluate the performance of convolution neural networks. Cross entropy function value increases if predicated value and actual value are not same. In ideal case, cross entropy value is zero.

In our instance, the minimum value of cross entropy is 0.0276 after 145 epochs. It seems to be nearby zero. In our case maximum error comes out to be 0.0562, which is combined form of training and validation. The training and validation accuracy of our convolution neural network for Kaggle dataset is also shown in Figure 4. The maximum value of training accuracy comes out to be 0.9905 after 145 epochs and the maximum value of validation accuracy is 0.9822.

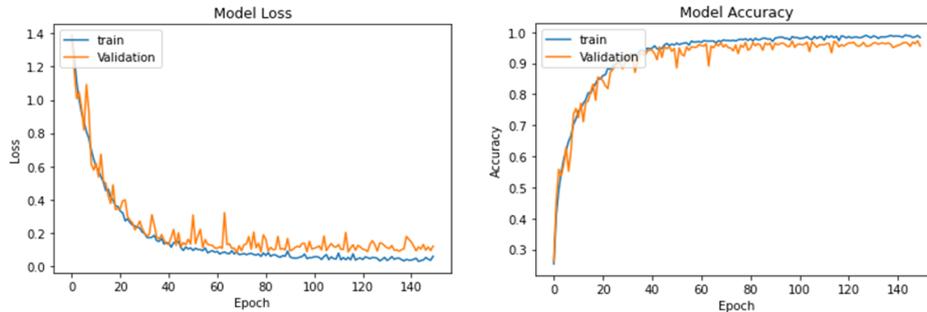

Fig. 4. Graphs representing model loss and accuracy of proposed CNN model on Kaggle dataset

Table 2. Confusion matrix of Kaggle dataset

| Class | EOSINOPHIL | LYMPHOCYTE | MONOCYTE | NEUTROPHIL |
|---|---|---|---|---|
| EOSINOPHIL | **246** | 0 | 0 | 2 |
| LYMPHOCYTE | 0 | **248** | 0 | 0 |
| MONOCYTE | 0 | 0 | **248** | 0 |
| NEUTROPHIL | 1 | 0 | 0 | **247** |

Table 3. Test results of proposed CNN architecture

| Dataset | Type | Truth | Classified | Accuracy (%) | Precision (%) | Recall (%) | Fmeasure (%) |
|---|---|---|---|---|---|---|---|
| Kaggle | Eosinophils | 246 | 248 | 0.990 | 0.99 | 0.994 | 0.993 |
|  | Lymphocytes | 248 | 248 | 100 | 100 | 100 | 100 |
|  | Monocytes | 248 | 248 | 100 | 100 | 100 | 100 |
|  | Neutrophils | 247 | 248 | 0.993 | 0.993 | 0.985 | 0.98 |
| LISC | Eosinophils | 97 | 99 | 0.985 | 0.985 | 0.998 | 0.98 |
|  | Lymphocytes | 96 | 99 | 0.972 | 0.972 | 0.952 | 0.977 |
|  | Monocytes | 99 | 99 | 100 | 100 | 100 | 100 |
|  | Neutrophils | 98 | 99 | 0.99 | 0.99 | 0.99 | 0.98 |

With the proposed CNN model, model weights for minimum loss are saved and labels for testing dataset are predicted. Results in the form of confusion matrix are presented in Table 2. For Kaggle dataset, 2 images of eosinophils and 1 image of neutrophils class are classified incorrectly. It misclassifies the Eosinophils and Neutrophils because as explained earlier, the two types of cells are similar in shape and size. All images of other two classes are 100% correctly classified. From Table 3 reported next, it can be seen that for Eosinophils, Neutrophils, Lymphocytes, Monocytes class, the precision is 99%, 99.3%,100% and 100% respectively. The F-measure rates of the four classes separately obtained are 99.3% in Eosinophils, 98%, in Neutrophils, 100% in Lymphocytes and 100% in Monocytes. For the Eosinophils, Neutrophils, Lymphocytes, Monocytes class, the recall is 99.4%, 98.5%, 100% and 100% respectively. It mixed up the Eosinophils and Neutrophils, because they are near identical and have same sizes and shapes. All images of Lymphocytes and Monocytes are 100% correctly classified. Average accuracy achieved for all classes is 99.57%.

**Results on LISC Dataset.** Graph for model loss of LISC dataset and accuracy graph is shown in Figure 5. Here, the minimum value of cross entropy loss function is 0.0354 after 147 epochs. The maximum value of validation accuracy comes out to be 0.983 and the maximum value of training accuracy is 0.9712.

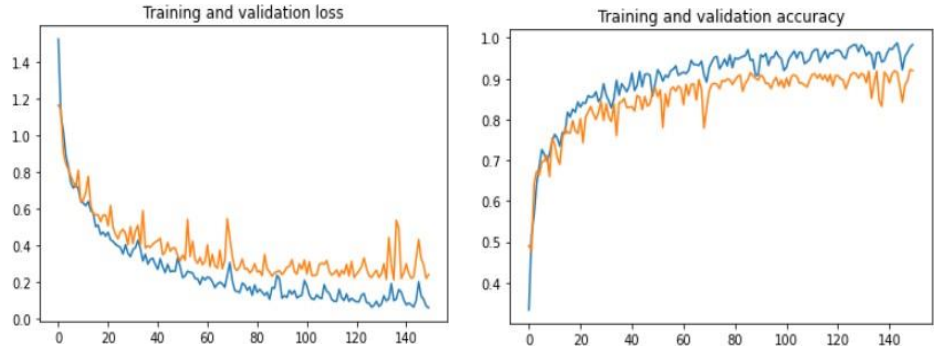

Fig. 5. Graphs representing model loss and accuracy of proposed CNN model on LISC dataset

**Table 4.** Confusion matrix of LISC dataset

| Class | EOSINOPHIL | LYMPHOCYTE | MONOCYTE | NEUTROPHIL |
|---|---|---|---|---|
| EOSINOPHIL | **97** | 0 | 0 | 2 |
| LYMPHOCYTE | 1 | **96** | 0 | 2 |
| MONOCYTE | 0 | 0 | **99** | 0 |
| NEUTROPHIL | 1 | 0 | 0 | **98** |

Confusion matrix is shown in Table 4 for LISC dataset. 2 images of eosinophils, 3 of lymphocytes and 1 of neutrophils, hence 6 images in total of 396 are misclassified. Average accuracy of 98.67% is achieved. Performance parameters of white blood cells classification obtained with proposed CNN model for both datasets are again shown in Table 3.

We compared the results of our proposed method with other related works from literature given in Table 5. To our knowledge, Ergen et. al. [14] achieved the accuracy of 97.95% using CNN on Kaggle dataset. Using pre-trained deep learning models, Mohamed et. al. [17] reported an accuracy of 97.03% on BCCD dataset. Accuracy of 96.63% has been achieved on ALL-IDB dataset by Macawile et al. [6] using AlexNet network. Our results on the two datasets are competitive as compared to results reported by Cinar et al [20] as they have also employed the same two datasets. Using Alexnet-GoogleNet-SVM, Cinar et. al. [20] achieved an accuracy of 99.73% for Kaggle dataset and 98.23% for LISC dataset. Our Convolutional Neural Network model has performed very well for the categorization of WBCs giving an accuracy of 99.57% for Kaggle dataset and 98.67% for LISC dataset. This shows that our proposed CNN model is effective in correctly classifying the 4 types of WBCs and competes with results reported in literature on the two datasets.

Table 5. Comparison with related work

| Author | Year | Dataset | Training/Testing Images | Method | Accuracy |
|---|---|---|---|---|---|
| M. J. Macawile [6] | 2018 | ALL-IDB dataset | N/A | AlexNet | 96.63% |
| G. Liang [9] | 2018 | BCCD, Kaggle | 12000 | CNN + RNN | 91% |
| M. Sharma[13] | 2019 | BCCD | 9957/2487 | CNN | 87.93% |
| M. Togacar [15] | 2019 | Local Dataset | N/A | CNN | 97.78% |
| E.H. Mohamed [17] | 2020 | BCCD | 2500/620 | Pre-trained Deep Learning Models | 97.03% |
| B. Ergen [14] | 2020 | Kaggle | 8710/3733 | CNN, Feature Selection | 97.95% |
| C. Zhao [24] | 2021 | BCCD | N/A | TWO-DCNN | 96% |
| A. Cinar [20] | 2021 | Kaggle, LISC | N/A | Alexnet-GoogleNet-SVM | 99.73%, 98.23% |
| **Proposed Method** | **2022** | **Kaggle Dataset, LISC Dataset** | **12500, 10,000** | **CNN** | **99.57%, 98.67%** |

## 4 Conclusion

In this paper, we have first applied different pre-trained models, InceptionV3, VGG16, MobileNetV2 and ResNet50. Motivated by these architectures and a quest for better performance, a CNN based model has been proposed to categorize the sub types of white blood cells (WBCs). The proposed architecture consists of three convolutional layers, three pooling layers, two fully connected hidden layers and output layer. The classification is performed using microscopic blood cell images obtained from Kaggle and LISC dataset. During testing, the proposed algorithm has shown optimal performance in terms of classification with 99.57% accuracy for Kaggle dataset and 98.67% for LISC dataset. The proposed model is effective as the results achieved are competitive in comparison with previous results reported in literature on the same datasets. In future, our proposed architecture can be applied to the classification of other cells and tissues of body that can help the pathologists in effective diagnosis.

## References

1. M. Z. Alom, C. Yakopcic, T. M. Taha, and V. K. Asari, "Microscopic Blood Cell Classification Using Inception Recurrent Residual Convolutional Neural Networks," in *NAECON 2018 - IEEE National Aerospace and Electronics Conference*, Dayton, OH, Jul. 2018, pp. 222–227.
2. R. Ahasan, A. U. Ratul, and A. S. M. Bakibillah, "White Blood Cells Nucleus Segmentation from Microscopic Images of strained peripheral blood film during Leukemia and Normal Condition," in *2016 5th International Conference on Informatics, Electronics and Vision (ICIEV),* Dhaka, Bangladesh, May. 2016, pp. 361-366.
3. Puttamadegowda J. and Prasannakumar S. C., "White Blood cell sementation using Fuzzy C means and snake," in *2016 International Conference on Computation System and Information Technology for Sustainable Solutions (CSITSS)*, Bengaluru, India, Oct. 2016, pp. 47–52.

4. S. Manik, L. M. Saini, and N. Vadera, "Counting and classification of white blood cell using Artificial Neural Network (ANN)," in *2016 IEEE 1st International Conference on Power Electronics, Intelligent Control and Energy Systems (ICPEICES)*, Delhi, India, Jul. 2016, pp. 1–5.
5. T. Rosyadi, A. Arif, Nopriadi, B. Achmad, and Faridah, "Classification of leukocyte images using K-Means Clustering based on geometry features," in *2016 6th International Annual Engineering Seminar (InAES)*, Yogyakarta, Indonesia, Aug. 2016, pp. 245–249.
6. M. J. Macawile, V. V. Quinones, A. Ballado, J. D. Cruz, and M. V. Caya, "White blood cell classification and counting using convolutional neural network," in *2018 3rd International Conference on Control and Robotics Engineering (ICCRE)*, Nagoya, Apr. 2018, pp. 259–263.
7. Gautam, P. Singh, B. Raman, and H. Bhadauria, "Automatic classification of leukocytes using morphological features and Naïve Bayes classifier," *IEEE Reg. 10 Annu. Int. Conf. Proceedings/TENCON*, pp. 1023–1027, 2017.
8. W. Yu, J. Chang, C. Yang, L. Zhang, H. Shen, Y. Xia, and J. Sha, "Automatic classification of leukocytes using deep neural network," in *2017 IEEE 12th International Conference on ASIC (ASICON)*, Guiyang, Oct. 2017, pp. 1041–1044.
9. G. Liang, H. Hong, W. Xie, and L. Zheng, "Combining Convolutional Neural Network With Recursive Neural Network for Blood Cell Image Classification," *IEEE Access*, vol. 6, pp. 36188–36197, 2018.
10. P. Tiwari, J. Qian, Q. Li, B. Wang, D. Gupta, A. Khanna, and J.J.P.C. Rodrigues, "Detection of subtype blood cells using deep learning," *Cogn. Syst. Res.*, vol. 52, pp. 1036–1044, Dec. 2018.
11. M. Z., T. S., and A. B., "Neural Network Classification of White Blood Cell using Microscopic Images," *Int. J. Adv. Comput. Sci. Appl.*, vol. 8, no. 5, 2017, pp. 99-104.
12. H. Fan, F. Zhang, L. Xi, Z. Li, G. Liu, and Y. Xu, "LeukocyteMask: An automated localization and segmentation method for leukocyte in blood smear images using deep neural networks," *J. Biophotonics*, vol. 12, no. 7, Jul. 2019.
13. M. Sharma, A. Bhave, and R. R. Janghel, "White Blood Cell Classification Using Convolutional Neural Network," in *Soft Computing and Signal Processing*, vol. 900, J. Wang, G. R. M. Reddy, V. K. Prasad, and V. S. Reddy, Eds. Singapore: Springer Singapore, 2019, pp. 135–143.
14. M. Toğaçar, B. Ergen, and Z. Cömert, "Classification of white blood cells using deep features obtained from Convolutional Neural Network models based on the combination of feature selection methods," *Appl. Soft Comput.*, vol. 97, p. 106810, Dec. 2020.
15. M. Togacar, B. Ergen, and M. E. Sertkaya, "Subclass Separation of White Blood Cell Images Using Convolutional Neural Network Models," *Elektron. Ir Elektrotechnika*, vol. 25, no. 5, pp. 63–68, Oct.2019.
16. I. Journal, X. Yao, K. Sun, X. Bu, C. Zhao, and Y. Jin, "Classification of white blood cells using weighted optimized deformable convolutional neural networks convolutional neural networks," *Artif. Cells, Nanomedicine, Biotechnol.*, vol. 49, no. 1, pp. 147–155, 2021.
17. E. H. Mohamed, W. H. El-Behaidy, G. Khoriba, and J. Li, "Improved White Blood Cells Classification based on Pre-trained Deep Learning Models," *J. Commun. Softw. Syst.*, vol. 16, no. 1, pp. 37–45, Mar. 2020.
18. https:// www. kaggle.com/ paultimothymooney/ blood- cells
19. http://users.cecs.anu.edu.au/~hrezatofighi/Data/Leukocyte%20Data.htm
20. Çınar and S. A. Tuncer, "Classification of lymphocytes, monocytes, eosinophils, and neutrophils on white blood cells using hybrid Alexnet-GoogleNet-SVM," *SN Appl. Sci.*, vol. 3, no. 4, p. 503, Apr. 2021.
21. C. Szegedy, V. Vanhoucke, S. Ioffe, J. Shlens, and Z. Wojna, "Rethinking the Inception Architecture for Computer Vision," in *2016 IEEE Conference on Computer Vision and Pattern Recognition (CVPR)* , Las Vegas, NV, USA, Jun. 2016, pp. 2818–2826.
22. K. He, X. Zhang, S. Ren, and J. Sun, "Deep Residual Learning for Image Recognition," in

*2016 IEEE Conference on Computer Vision and Pattern Recognition (CVPR)*, Las Vegas, NV, USA, Jun. 2016, pp. 770–778.
23. M. Sandler, A. Howard, M. Zhu, A. Zhmoginov, and L.-C. Chen, "MobileNetV2: Inverted Residuals and Linear Bottlenecks," in *2018 IEEE/CVF Conference on Computer Vision and Pattern Recognition*, Salt Lake City, UT, Jun. 2018, pp. 4510–4520.
24. I. Journal, X. Yao, K. Sun, X. Bu, C. Zhao, and Y. Jin, "Classification of white blood cells usingweighted optimized deformable convolutional neural networks convolutional neural networks,"*Artif.Cells,Nanomedicine,Biotechnol.*,vol.49, no.1, pp.147–155, 2021.